\documentclass[twocolumn]{aastex631}

\usepackage{multirow}
\usepackage{comment}   

\shorttitle{AASTeX v6.3.1 Sample article}
\shortauthors{Duggal et al.}
\graphicspath{{./}{figures/}}

\begin{document}

\title{Mapping feedback signatures in 3C 297: A quasar-host merger at Cosmic Noon}


\correspondingauthor{Chetna Duggal}
\email{duggalc@myumanitoba.ca}

\author[0000-0001-7781-246X]{C. Duggal}
\affiliation{Department of Physics and Astronomy, University of Manitoba, Winnipeg, MB R3T 2N2, Canada} 

\author[0000-0001-6421-054X]{C. P. O'Dea}
\affiliation{Department of Physics and Astronomy, University of Manitoba, Winnipeg, MB R3T 2N2, Canada}

\author[0000-0002-4735-8224]{S. A. Baum}
\affiliation{Department of Physics and Astronomy, University of Manitoba, Winnipeg, MB R3T 2N2, Canada}

\author{J. Jiwa}
\affiliation{Department of Physics and Astronomy, University of Manitoba, Winnipeg, MB R3T 2N2, Canada}

\author[0000-0002-5445-5401]{G. R. Tremblay}
\affiliation{Harvard-Smithsonian Center for Astrophysics, 60 Garden Street, Cambridge, MA 02138, USA}

\author{M. Chiaberge} 
\affiliation{Space Telescope Science Institute, 3700 San Martin Drive, Baltimore, MD 21218, USA}

\author{G. Miley} 
\affiliation{Leiden Observatory, University of Leiden, P.O. Box 9513, Leiden, 2300 RA, The Netherlands}

\author[0000-0002-6415-854X]{C. Stanghellini} 
\affiliation{INAF-Istituto di Radioastronomia, Via Gobetti 101, I-40129 Bologna, Italy}

\author{W. Sparks} 
\affiliation{SETI Institute, 339 N Bernardo Avenue, Mountain View, CA 94043, USA}



\begin{abstract}
We present a study of quasar host galaxy 3C 297 which is home to a powerful bent-jet radio source suggesting vigorous interaction with a dense ISM and/or jet precession. Archival HST imaging showed interestingly perturbed morphology of the host with a bright $\sim$30 kpc arc feature, extended filamentary structure of line-emitting gas and clumpy blue excess emission co-spatial with the radio hotspots. Our VLT/SINFONI integral-field observations reveal complex, spatially-resolved H$\alpha+$[NII] emission in this source. A prominent blue-shifted wing in H$\alpha$ indicates an ionized gas flow extending out to $\sim$18 kpc from the nuclear region. Combining our SINFONI narrow-H$\alpha$ data with archival HST/UV and VLA imaging, we map the young stellar population in the host and compare the spatial distribution of star-forming regions with the ionized gas motion and jet structure. In the attempt to characterize the feedback mechanisms in this chaotic system, we suggest that the powerful radio source dominates the feedback with possible contribution from radiation pressure due to AGN accretion. We also propose that the expanding jet cocoon likely shocked the ISM,  triggering a kpc-scale ionized gas outflow and new starbursts that enhanced ongoing merger-induced star formation.

\end{abstract}


\section{Introduction} \label{sec:intro}

The epoch between redshifts $1<z<3$ holds a significant place in cosmic history in the context of both black hole and galaxy evolution. Referred to as Cosmic Noon, this period represents a peak in star formation rate density (\citealt{2014ARA&A..52..415M, 2020ARA&A..58..661F}) as well as quasar number density (\citealt{1998MNRAS.293L..49B, 2000MNRAS.311..576K, 2008AJ....136.2373R, 2020MNRAS.495.3252S}). Given the well-known tight correlations between supermassive black hole (SMBH) and galaxy properties (e.g., \citealt{1998A&A...331L...1S, 2005ApJ...635L.121K, 2013ARA&A..51..511K}) and the fact that most massive galaxies had already formed by $z\sim1$ (e.g., \citealt{2006A&A...453L..29C}), it is crucial to probe the efficiency of energy feedback from accreting black hole engines (Active Galactic Nuclei; AGN) in the period around the $z=2$ peak of the cosmic black hole accretion rate. 
\newline

Quasar-host galaxies are commonly found to be enduring the impacts of powerful AGN feedback in the form of gaseous outflows. 
Fast, massive outflows have been observed in ionized, atomic, and molecular gas phases on distance scales from galactic ISM at $\sim$1 kpc up to $\sim$100 kpc circumgalactic halos (e.g., \citealt{2018MNRAS.479.5544M, 2020MNRAS.497.5229C, 2020MNRAS.491.4462D, 2023Galax..11...73B}) within the quasar hosts (see also the recent review by \cite{2024Galax..12...17H}). Outflows may be powered by radiation pressure from the accretion disk or by the mechanical action of the relativistic jet plasma.
How galaxy-wide outflows affect the interstellar medium (ISM) and star forming activity in the host, is an important facet of the AGN feedback paradigm and has been the focus of investigation in simulation and observational studies (e.g., 
\citealt{2003MNRAS.345..657K, 2005ARA&A..43..769V, 2013Sci...341.1082M, 2016ApJ...817..108W, 2018ApJ...865....5R, 2017A&A...601A.143F, 2018NatAs...2..176C, 2019ApJ...879...75H, 2021MNRAS.503.1780J, 2023A&A...678A.127V, 2024MNRAS.531.2079M}).
\newline

\begin{figure*}[ht!]
\centering
\epsscale{1.2}
\plotone{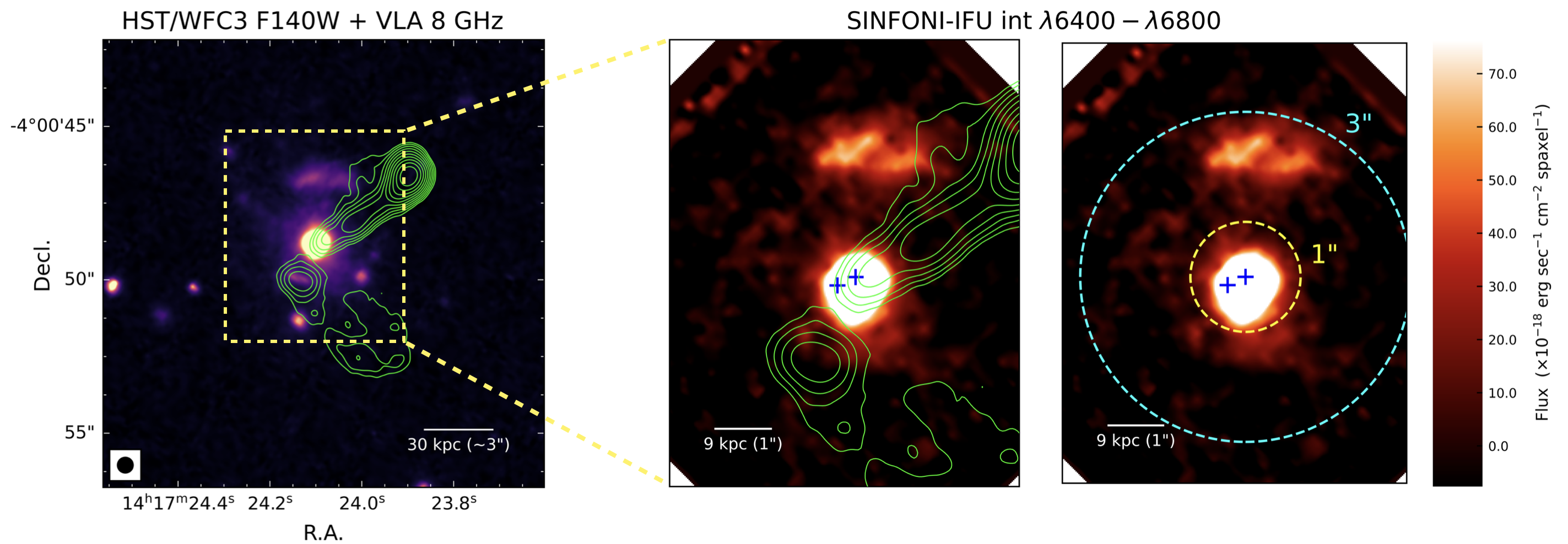}
\caption{HST/WFC3 image of 3C 297 in F140W filter with VLA 8.4 GHz radio contours (plotted at intervals defined by $2^i \times 3\sigma$ mJy beam$^{-1}$, where $i=1, 2, 3, ..., 10$) overlaid \emph{(left)}. The yellow dashed lines indicate the $8''\times8''$ SINFONI FOV. The middle and right panels show a 2D slice of the SINFONI datacube integrated over the H$\alpha$ line profile. Blue cross points mark the possible double nuclei positions and green contours depict the radio source structure \emph{(middle)}, and the dashed circles outline the integration apertures \emph{(right)}. North is up, East is to the left. \label{fig:first}}
\end{figure*}

Kiloparsec-scale ionized gas outflows have been resolved and studied in active galaxies at $z>1$ (e.g. \citealt{2012A&A...537L...8C, 2015A&A...580A.102C, 2015ApJ...799...82C, 2020A&A...642A.147K, 2023A&A...672A.128C, 2023ApJ...953...56V, 2024A&A...683A.169W, 2024ApJ...960..126V}). 
The exact outflow morphologies (e.g., conical or shell-like) and driving mechanisms are poorly constrained, as is how their energy couples with the galactic environment as they propagate through the host ISM to reach kiloparsec scales.
\newline

Galaxy mergers are thought to be important triggers for radio-loud AGN activity. This is supported by the findings of recent studies that show radio-luminous AGN overwhelmingly reside in ongoing or recent major galaxy mergers (e.g., \citealt{2012ApJ...758L..39T, 2015ApJ...806..147C, 2024ApJ...963...91B}). Moreover, the detection of dual AGN in a merging system offers unique opportunity to probe the relationship between radio AGN and their host galaxy environs, as well as the role of major mergers in triggering of AGN activity (e.g., \citealt{2024MNRAS.535..763R, 2023ApJ...951L..18G}).
\newline

At a redshift of $z$=1.4, 3C 297 is host to a powerful (L[178 MHz]$\sim$10$^{36}$ erg sec$^{-1}$ Hz$^{-1}$; \citealt{2016ApJS..225...12H}), high-excitation (deduced from optical line emission; \citealt{1997MNRAS.286..241J}) radio source. Thus, it is expected to have feedback contribution from the quasar itself, in addition to the kinetic jet-mode feedback, owing to the radiatively efficient accretion. 
It has been hypothesized that the host galaxy of 3C 297 is a fossil group, i.e. the stellar mass from other gravitatonally-bound neighbour galaxies has already merged into it, suggested by the presence of an X-ray-luminous halo and a lack of companion galaxies (\citealt{2023ApJS..264....6M}). 

3C 297 was chosen for integral-field follow-up observations from a sample of 58 high-redshift 3CR \citep{1985PASP...97..932S} sources imaged as part of the Hubble Space Telescope (HST) infrared snapshot survey \citep{2016ApJS..225...12H}. The HST/WFC3 imaging data cover rest-frame ultraviolet (UV; F606W) and optical (F140W) bands and explore 3C 297's highly perturbed morphology, showing extended line-emitting regions, elongated filament and arc-like structures as well as knots of UV-excess emission exhibiting remarkable spatial correlation with the radio source. 

In the F140W image (Figure \ref{fig:first}, left panel), bright optical emission is detected from an arc-shaped region, about $\sim$30 kpc North of the core. An extended H$\alpha$ filament is also observed, extending to $\sim$25 kpc towards the Northeast. The F606W image resolved the core of 3C 297 into two distinct blobs of UV emission, with the Eastern clump more extended than the compact Western nucleus (coinciding with the AGN position). The complex, disturbed morphology of 3C 297 and possible double nuclei form the basis of its identification as an ongoing merger \citep{2015ApJ...806..147C}.
\newline

Archival radio imaging from the Very Large Array (VLA) maps the high-power jet emission. The 8.4 GHz image (contour overlay in Figure \ref{fig:first}) shows $\sim$40 kpc FR II \citep{1974MNRAS.167P..31F} radio jets extending from the core. 
The Northern jet lobe is likely interacting with the bright line-emission arc feature which may be an ``exit wound" associated with gas swept aside as the radio source escapes the dense ISM;  while emission from the Southern jet is highly bent and spread over a larger area with two hotspots co-spatial with the elongated, clumpy emission in the UV. 

Deflection in a jet's path is expected when it interacts with a dense ISM, resulting in sharp bends in the observed radio morphology (e.g., \citealt{2000ApJ...534..201W, 2004A&A...424..119M, 2018MNRAS.479.5544M, 2025ApJ...981..149B}). This deviation could be more drastic in an active merger environment resulting in large bending angles (e.g., \citealt{2010MNRAS.407..721J}). 3C 297's Southern jet appears to bend at a $90^\circ$ angle at the site of the radio hotspots and may also have a possible $\approx$180$^\circ$ bend at the Southwest edge of the lobe. Moreover, presence of precessing jets will also cause more extended radio emission perpendicular to the jet axis, depending on the travel time of the jet from core to hotspot. Jet precession can be induced by the orbital motion of binary black holes in merging systems (\citealt{2019MNRAS.482..240K, 2025A&A...695A.179S}).

Both the ISM gas interaction and SMBH pair-driven precession may be responsible for the perturbed jet morphology in the 3C 297 system. Interestingly,  distant (redshift $>1.5$) quasars are known to distinctly exhibit more bent, distorted radio sources with smaller radio sizes than those at nearby redshifts (\citealt{1988A&AS...73..515B, 1988Natur.333..319B}). This epoch-dependent morphology is attributed to the interaction of jets with the denser ambient medium in younger galaxies.
\newline

With the aim of probing its chaotic host-AGN interaction dynamics, we obtained spatially-resolved spectroscopy and map the distribution and kinematics of ionized gas in 3C 297. Combining our 3D spectroscopic data from the Spectrograph for INtegral Field Observations in the Near Infrared (SINFONI) with archival imaging at the optical, infrared and radio wavelengths, we present here an analysis of AGN feedback signatures in this source and discuss their driving mechanisms.

Throughout this work, a flat $\Lambda$CDM cosmology with H$_0$ = 69.6 km s$^{-1}$ Mpc$^{-1}$, $\Omega_M$ = 0.3 and $\Omega_{vac}$ = 0.7 is assumed.

\section{Observations and data reduction} \label{sec:obs}

The galaxy 3C 297 was observed as part of the program 097.B-0452(A) (PI: G. Tremblay) with the SINFONI instrument mounted on the UT4 telescope at the Very Large Telescope (VLT) until 2019. The observations were executed on 13 March 2017, in closed-loop adaptive optics (AO) mode with an artificial sodium laser guide star, due to lack of a suitable natural AO guide star within range of the
target. The nucleus of 3C 297 served as
guide star for the tip-tilt correction. The ambient sky conditions were relatively clear and stable with average near-IR seeing $\sim0.8''$ between the start and end of the observation.

The data cover the H-band grating (1.45$-$1.85 $\mu$m) covering the redshifted H$\alpha$, [N II] and [S II] emission and providing spectral resolution of R = 3000. 
The data were reduced using the ESO-SINFONI pipeline version 2.9.0. After flat-fielding, sky subtraction, correction for distortions, cosmic rays removal and wavelength calibration, the final co-added data cube has a spatial scale of 0.125$''$ pixel$^{-1}$ and a total field of view of $8'' \times 8''$. The angular resolution, as sampled by the point-spread function (PSF) from the standard star observations, is $\sim$0.4$''$. At the target redshift, this translates to spatial resolution of $\sim$3 kpc at 8.57 kpc/$''$.
\newline

The flux scale of our observation was calibrated based on existing photometric data for this galaxy.  HST/F140W imaging observation from \cite{2016ApJS..225...12H} reported Galactic extinction-corrected magnitude of 19.377 $\pm$ 0.003, where it was pointed out that while the bandpass includes H$\alpha$ and [O III]$\lambda$5007 emission lines, [O III]$\lambda$5007 contributes to 0.03$\%$ of the total flux in the bandpass, as calculated from the \cite{1997MNRAS.286..241J} measurement. Since the continuum level is also very low compared to the H$\alpha$ line flux in the observed spectrum, we use the F140W flux to derive the scale factor for flux density of H$\alpha$ emission in the SINFONI observation.

\section{Analysis and Results} \label{sec:analysis}

\begin{figure*}
\epsscale{1.2}
\plotone{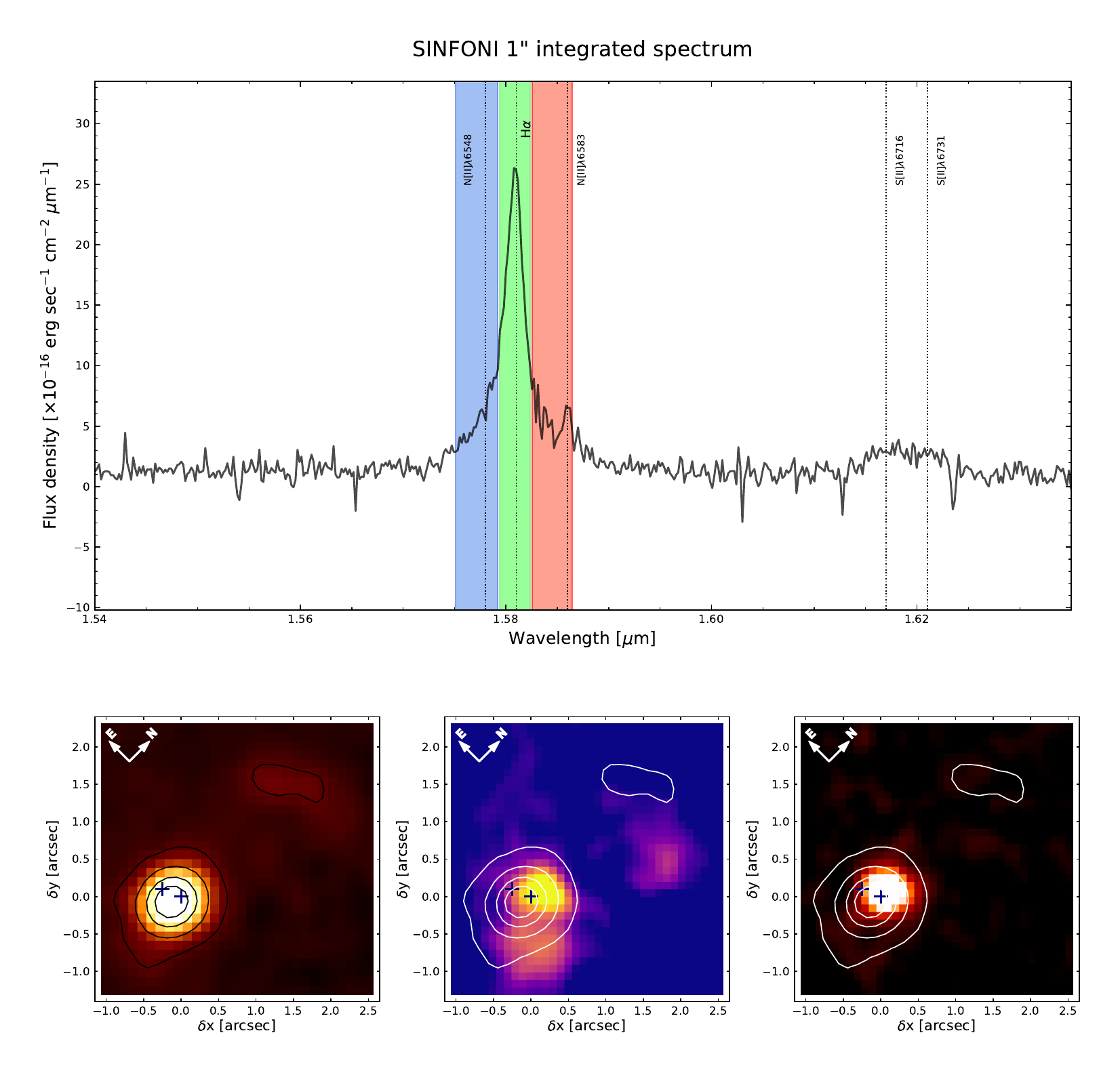}
\caption{\emph{(Top panel)} Observed SINFONI H-band spectrum of 3C 297, integrated by co-adding spaxels over an aperture of radius 1'' centered on the radio position. Gas emission lines detected in the spectrum are shown with dotted lines. The green, blue and red boxes show the wavelength intervals over which the maps shown in the bottom panels are integrated. \emph{(Bottom panels)} Emission maps obtained by collapsing the data cube on the line core ($1.579 < \lambda < 1.583 \mu$m, left, green shaded region), on the blue wing ($1.574 < \lambda < 1.579 \mu$m, middle, blue shaded region) and the red wing ($1.583 < \lambda < 1.586 \mu$m, right, red shaded region). In all three panels, contours on the line core (levels 0.05, 0.2, 0.4, 0.6 relative to the peak) and '+' markers at the positions of the possible double nuclei are shown. The blue wing extending to $1''$ (or $\sim$8.6 projected kiloparsecs) from the central spaxel is clearly resolved.} \label{fig:second}
\end{figure*}

The integrated H-band spectrum of 3C 297 is shown in Figure \ref{fig:second}. De-blending stellar emission from the host prior to ionized-gas emission line analysis was not necessary as the spectrum is free of continuum emission from the host galaxy. This is in agreement with GMOS observations of \cite{2023ApJS..264....6M}. From the positions of the different lines, we measured a mean spectroscopic redshift of z = 1.409 $\pm$ 0.001.

\subsection{Emission-line measurements}\label{subsec:line}

The continuum level was estimated over adjacent line-free spectral intervals and subtracted from the cubes. Nuclear spectrum was extracted from the median-subtracted datacube for two apertures centered on the 
AGN position.  The 3'' aperture covers the entire source (radius 3''$\sim$13 kpc) and corresponds to the single-slit measurements of \cite{1997MNRAS.286..241J}. The 1'' aperture (radius 1''$\sim$9 kpc) focuses on the region around the nucleus that exhibits broad dispersion in H$\alpha$. This aperture is comparable with the single-slit measurements of \cite{2023ApJS..264....6M} in the rest wavelength range 2500$-$4500 \AA. 
\newline

Line fitting for the extracted spectra was done using the \textsc{scipy.leastsq}\footnote{https://docs.scipy.org/doc/scipy/reference/generated/scipy.optimize.leastsq.html} package in Python, which implements the Levenberg-Marquardt algorithm for non-linear least squares fitting. We model the H$\alpha$+[N II]$\lambda\lambda$6548,6583 complex along with the [S II]$\lambda\lambda$6716,6731 doublet with a multi-component Gaussian function, i.e., fitting each of the five emission lines simultaneously. 
The best fit model comprised of a narrow component for all five lines and a broad Gaussian component for H$\alpha$. The fit was constrained with the following conditions$-$ the relative centroid velocities of the H$\alpha$, [N II] and [S II] lines were fixed for each component and the same width was adopted for all narrow lines. Thus, all narrow lines were kinematically coupled to have the same line dispersion. The [N II] $\lambda$6548/$\lambda$6584 ratio was fixed at 0.32 \citep{2023AdSpR..71.1219D}. The free parameters were thus the broad H$\alpha$ centroid velocity and width, the narrow [N II] $\lambda$6584/H$\alpha$ and [S II] $\lambda$6716/H$\alpha$ flux ratio and the [S II] $\lambda$6716/$\lambda$6731 flux ratio. Uncertainties in the fit were computed using Monte Carlo simulations in which the emission lines are refit 100 times as the spectrum is perturbed according to an error spectrum constructed from noise in each spaxel.

\begin{deluxetable}{lrr}
\tablecaption{3C 297 emission-line fluxes \label{tab:flux}}
\tabletypesize{\footnotesize}
\tablehead{
\colhead{Line} & \colhead{1'' aperture} & \colhead{3'' aperture} \\
  &  \multicolumn{2}{c}{($10^{-14}$ erg s$^{-1}$ cm$^{-2}$)}
}
\startdata
\\
$\textrm{[O III]}\lambda$5007 &  \nodata   &   $^a$1.03  $\pm$  0.03  \\
$\textrm{[N II]}\lambda$6548 &   0.26  $\pm$  0.06   &   0.75 $\pm$  0.03\\
H$\alpha$          &  6.90  $\pm$  0.04 &    $^b$13.81 $\pm$  0.19 \\
$\textrm{[N II]}\lambda$6583 &  0.79  $\pm$  0.03  &   2.25  $\pm$  0.08  \\
$\textrm{[S II]}\lambda$6716 &   1.18  $\pm$  0.05 &   1.42  $\pm$  0.05  \\
$\textrm{[S II]}\lambda$6731 &   1.00  $\pm$  0.05  &    1.40  $\pm$  0.05 \\
\\
\enddata
\tablecomments{Galactic extinction-corrected fluxes derived as described in Sec. \ref{sec:obs}. $^a$\cite{1997MNRAS.286..241J}; $^b$\cite{2016ApJS..225...12H}}
\end{deluxetable}

The measured line velocities were corrected for SINFONI instrumental broadening. We used the average profile of three telluric OH lines to characterize the effective spectral resolution of our H-band data, as described by \cite{2009ApJ...706.1364F, 2018ApJS..238...21F}. This was done by extracting the night-sky emission spectrum in the 1$''$ aperture in the ``sky cube'' obtained by reducing the data once again but this time without background subtraction. The empirical line-spread function is well approximated by a Gaussian profile and fits give an effective spectral resolution corresponding to a velocity full width at half maximum (FWHM) of about 38 km s$^{-1}$ across the H-band.

Table \ref{tab:flux} lists the observed line fluxes integrated in 1" and 3" apertures. The integrated nuclear spectrum as well as the spatial map of the detected gas flow signature are illustrated in Figure \ref{fig:second}.

3C 297 shows a highly complex H$\alpha$+[N II] profile consistent with its perturbed morphology. A prominent blue wing is clearly visible in the H$\alpha$ emission line indicating the presence of outflowing material toward the observer. The fully resolved, extended blue wing is extended up to 1$''$, i.e., a projected distance of $\sim$8.6 kpc from the center towards the South-West. 
Our fitting yields the broad component FWHM = 2134 $\pm$ 545 km s$^{-1}$ and narrow component FWHM = 513 $\pm$ 57 km s$^{-1}$ in the galaxy core. The broad component shows a large velocity gradient with respect to the galaxy rest frame; the ionized gas velocity goes from a significant blueshift to an almost equal redshift within the inner 15 kpc. In general, we detect two unusual observations in tracing the ionized gas in this galaxy: (i) the spatial offset of outflow and broad-line components$-$ the South West region, while exhibiting a kpc-scale extended H$\alpha$ outflow, shows no broad component in the H$\alpha$ line; (ii) the entire arc-shaped extended line-emission region to the North is blue-shifted with respect to the core. Investigation of the nature and origin of these two features drives the analyses of this paper.

\subsection{Ionized gas distribution and kinematics}\label{broad}

\begin{figure*}
\epsscale{1.2}
\plotone{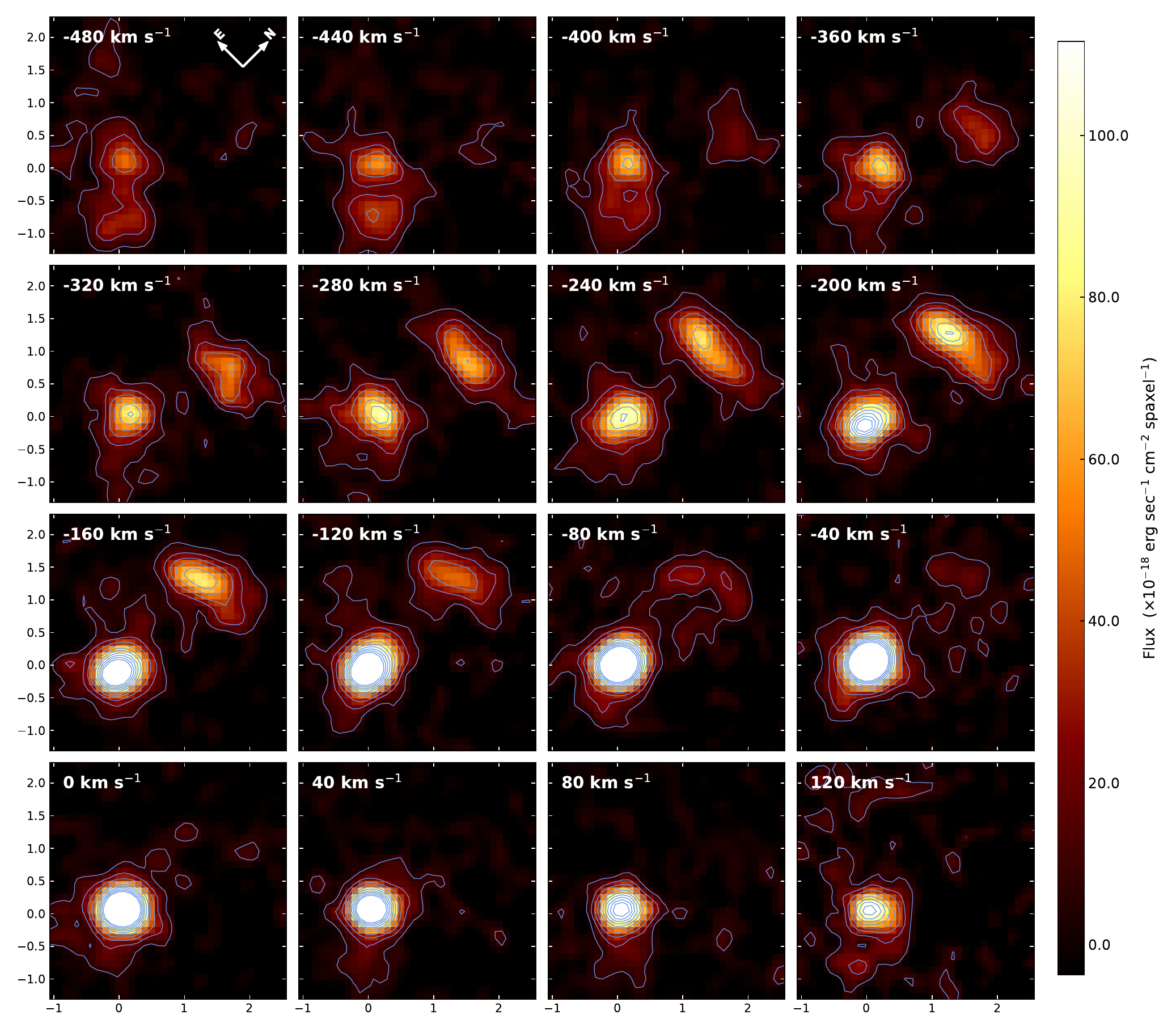}
\caption{SINFONI H-band channel maps showing 40 km s$^{-1}$ “slices” of the datacube from $-$480 km s$^{-1}$ through +120 km s$^{-1}$ relative to 3C 297 rest frame. The blue contours show the significance of the emission in multiples of $3\sigma$ increasing inwards towards the center. \label{fig:channel}}
\end{figure*}

Figure \ref{fig:channel} shows the channel maps covering the H$\alpha$+[N II] profile in 40 km s$^{-1}$ “slices” of the SINFONI datacube from $-$480 km s$^{-1}$ through +120 km s$^{-1}$ relative to the galaxy’s rest frame. 
The blue-shifted channels reveal the extended outflow of ionized gas, which appears as a one-sided flow from the core towards the South-West, with gas velocities of 250$-$400 km s$^{-1}$. It is possible that the detected outflow is one section of a larger, bi-directional outflow, with the red-shifted opposite direction hidden from the observer by dust.

The blue channels also reveal the nature of the Northern arc, which is likely a site of mechanical interaction of gas with the jet lobe as well as vigourous star formation. This is evident from the UV-bright clumps lining the entire arc (as detected in HST/F606W band \citep{2016ApJS..225...12H}; see Sec. \ref{subsec:sfr}), where the young stars power the H$\alpha$-emitting nebulae. 
\newline

\begin{figure*}
\centering
\epsscale{1.15}
\plotone{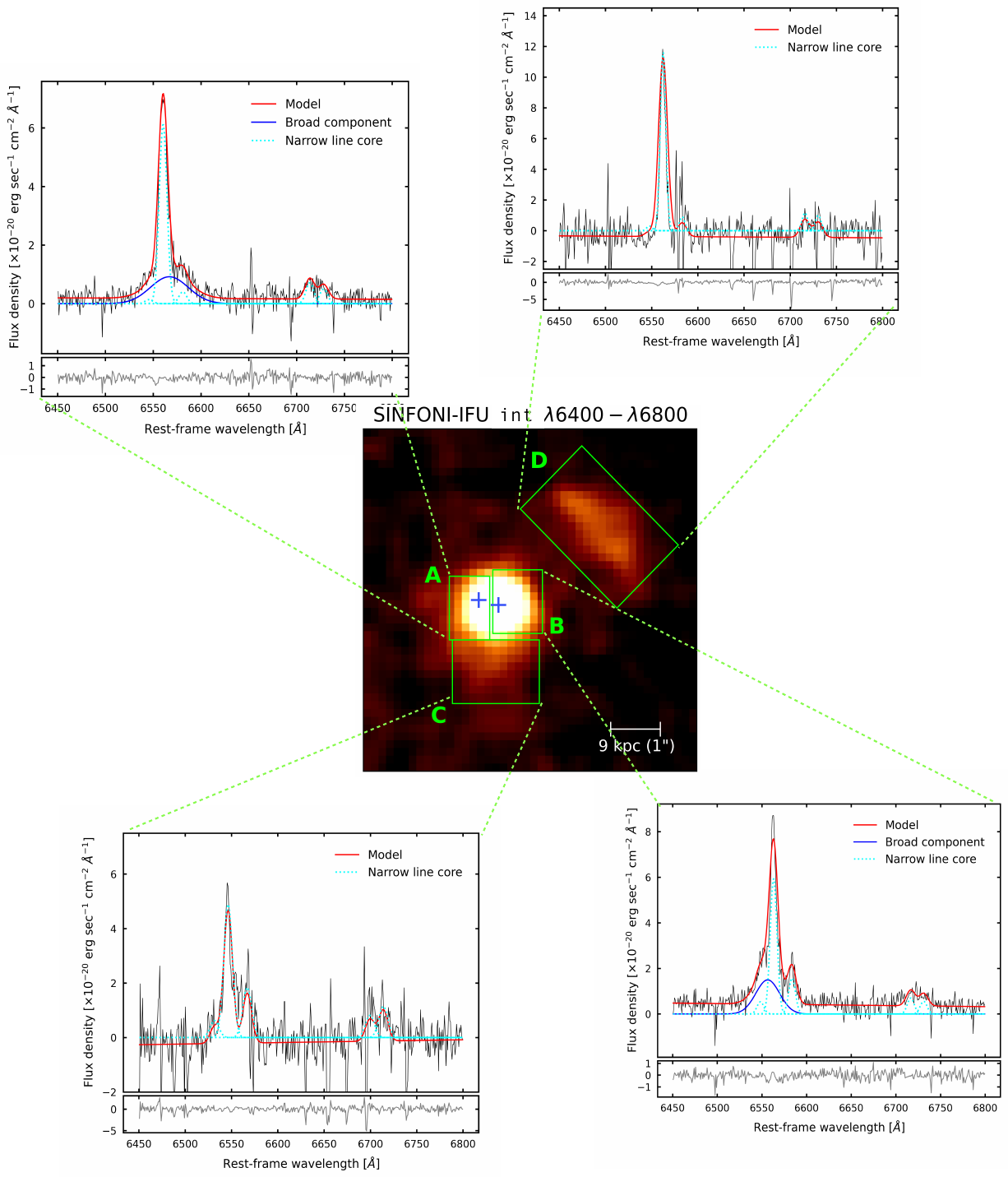}
\caption{Line velocities and velocity dispersions are derived from multi-Gaussian line fit (described in Sec \ref{subsec:line}) to the co-added spectrum from all the spaxels within each region (shown in solid green)}.  \label{fig:regions}
\end{figure*}

Figure \ref{fig:regions} dissects the ionized gas kinematics in 3C 297 region-wise, showing the narrow-line and broad-line characteristics on a slice of the SINFONI datacube integrated over the wavelength range $6400<\lambda$(\AA)$<6800$ that encompasses all the detected line emission. The region-specific emission-line gas properties derived from co-added spectra from the spaxels in each region are also listed. 

\begin{figure*}
\epsscale{1.1}
\plotone{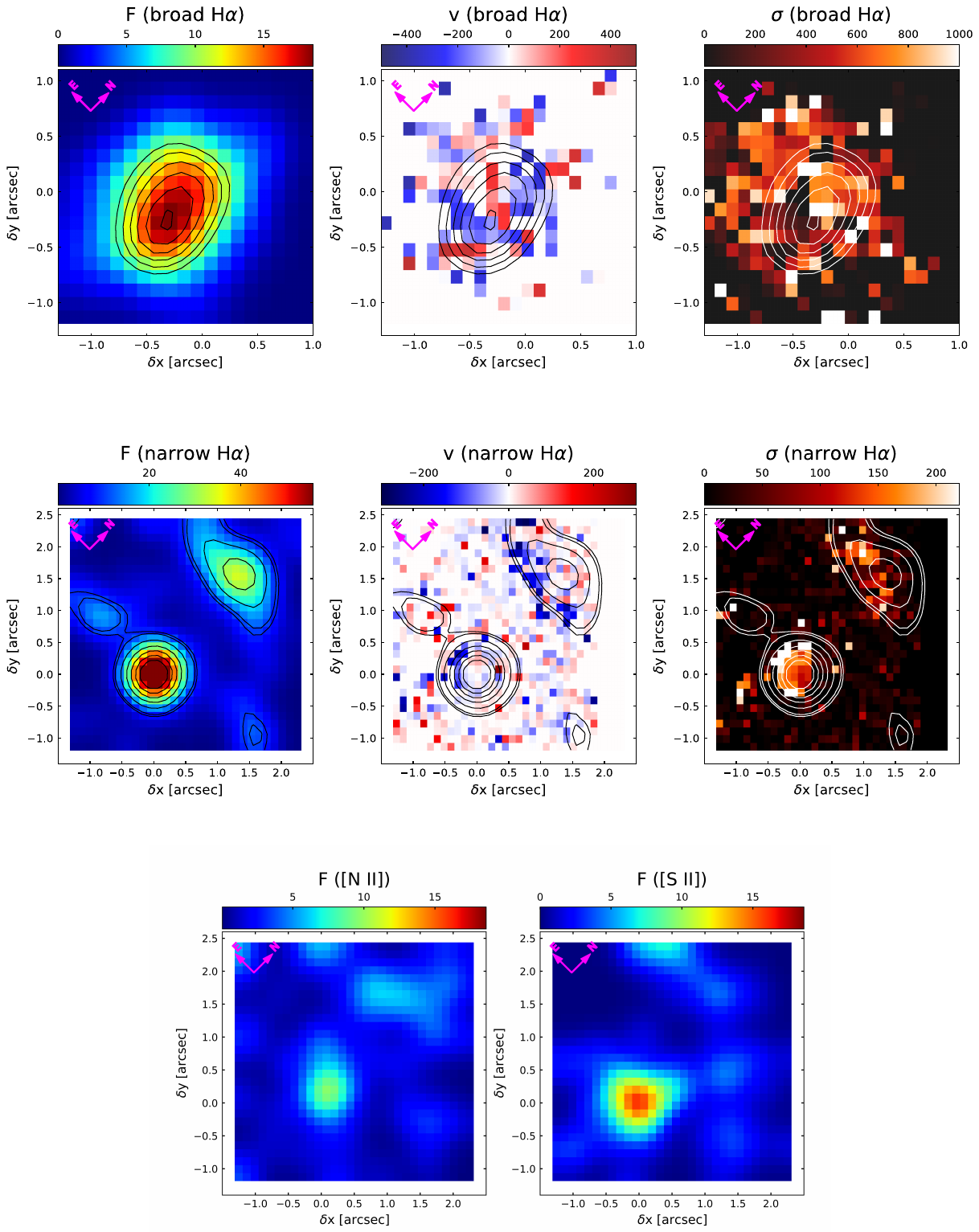}
\caption{Integrated line flux, velocity and velocity dispersion (right panels) distributions resulting from the coupled-Gaussian fitting, for each of the emission lines detected in the SINFONI H-band spectrum. The contour levels show the significance of emission (3$\sigma$ and above) in the flux maps, and are overlaid on velocity and velocity dispersion maps for spatial reference. All maps are centered on the radio position. Flux units
are $10^{-18}$ erg s$^{-1}$ cm$^{-2}$ spaxel$^{-1}$. Line velocity and velocity dispersion units are km s$^{-1}$. 
\label{fig:maps}}
\end{figure*}

The data in nuclear regions A and B show the broadest velocity distribution of ionized gas in the galaxy within the inner 15 kpc ($\sim$1.5$''$).
The fast-moving H$\alpha$ emitting gas exhibits a clear blue wing in region B which moves to the redder side in region A. This is observable with much better spatial clarity in Figure \ref{fig:maps}. The broad-line gas velocity map shows blocks of rotating gas$-$ gas clouds moving blueward on the outer edges and into the line-of-sight plane in the center$-$ perpendicular to the North East-South West direction. With higher resolution data, it would be possible to trace the rotation curves of the ionized gas and estimate the depth of the gravitational potential well in the core.

Region C covers the Southwestern extended ionized gas region which only shows narrow-line emission. The clearly resolved blue-shifted outflow structure (Figure \ref{fig:channel}) combined with the lower velocity dispersion in this region ($\sigma \sim$ 356 $\pm$ 31 km s$^{-1}$) suggests the bulk of the outflowing gas has slowed down after reaching a few kpc from the nucleus, toward the observer.

The direction of the observed gas flow is worth noting here. Gas outflows driven by AGN activity would be expected to be accelerated outwards along the jet axis or in the direction of the ionization cone formed by radiation from the accretion disk. In 3C 297, however, ionized gas motion appears to be in a direction that is not in the path of direct AGN radiation or jet plasma. It is possible that this feature may be gas flowing into the merging system from the surrounding intergalactic medium. Such an inflow of gas will move faster as it falls towards the nucleus, creating an increasing velocity gradient towards core region as observed, and will become ionized due to the expanding shock front of the jet cocoon. 

\begin{figure*}
\epsscale{1.2}
\plotone{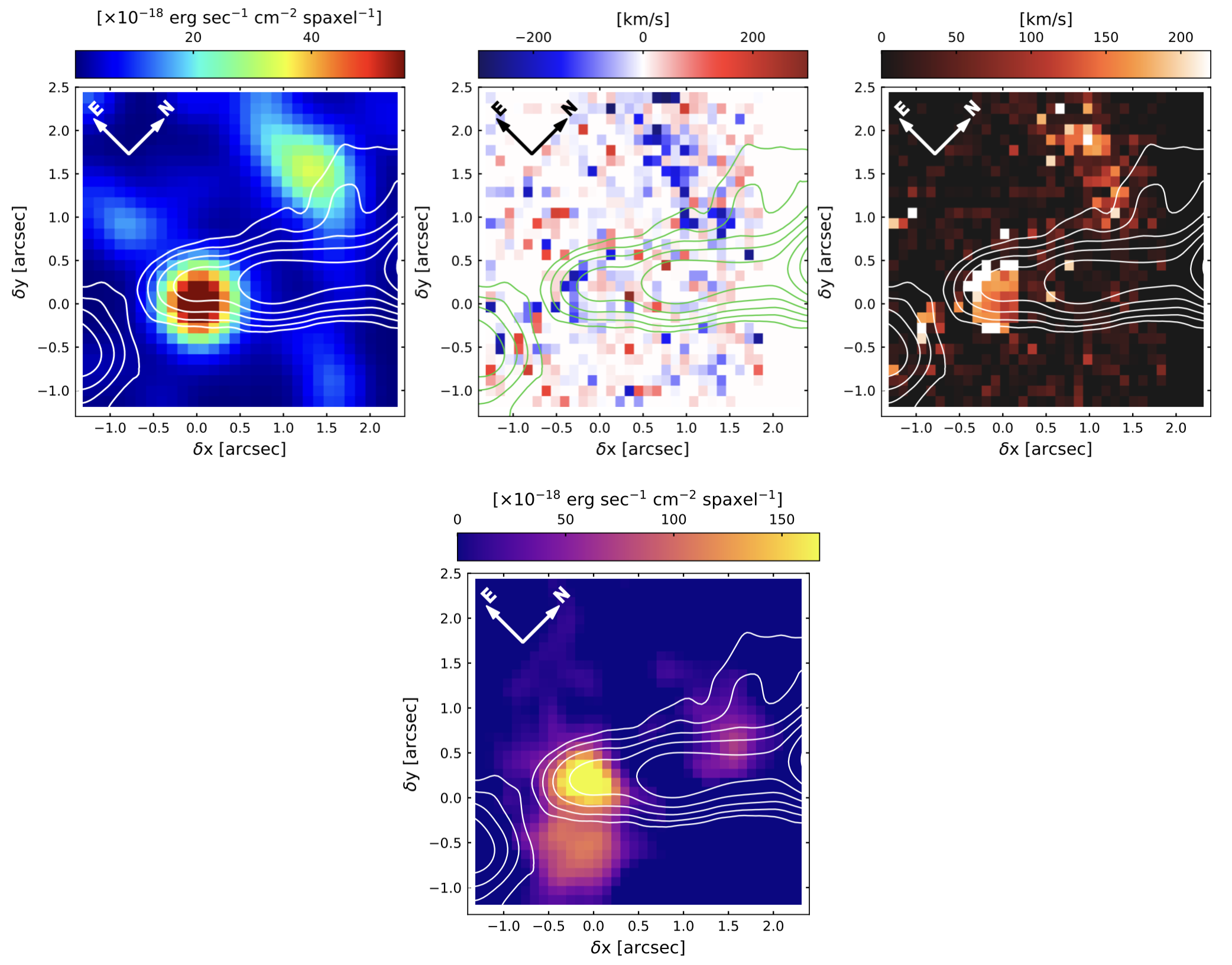}
\caption{Jet morphology with respect to H$\alpha$ distribution in 3C 297. \emph{Top panel:} H$\alpha$ flux  \emph{(left)}, narrow-H$\alpha$ velocity \emph{(middle)} and velocity dispersion \emph{(right)} maps. \emph{Bottom panel:}
The H$\alpha$ blue-wing map from Figure \ref{fig:second}. 
The contours from the 8.4 GHz radio image are overlaid on all maps. A clear spatial correlation is observed between the jet lobe with star formation-tracing H$\alpha$ emission within the Northern arc, as well as between the Southern hotspot of the jet and the blue-shifted gas flow. \label{fig:radio-overlays}}
\end{figure*}

Region D encompasses the Northern arc structure which shows only narrow ($\sigma \sim$ 126 $\pm$ 26 km s$^{-1}$) line emission. The H$\alpha$-bright region is likely lined with ionized gas nebulae being powered by star formation resulting from merger-related interaction in the outer rim of the host galaxy. Jet-driven shocks may also be a possible source of ionization in this region, as the bent lobe of the northern jet closely envelops the line-emitting gas arc. The velocity structure of ionized gas along the jet is shown in Figure \ref{fig:radio-overlays}. We investigate the sources of ionization in the following section. It is interesting to note here that shocks from the advancing jet lobe are likely also causing compressing dense clouds in this region, thus boosting the formation of the young stars that can ionize the line-emitting gas. We further discuss this scenario in Sec. \ref{subsec:sfr}.

\begin{figure*}
\centering
\epsscale{1.2}
\plotone{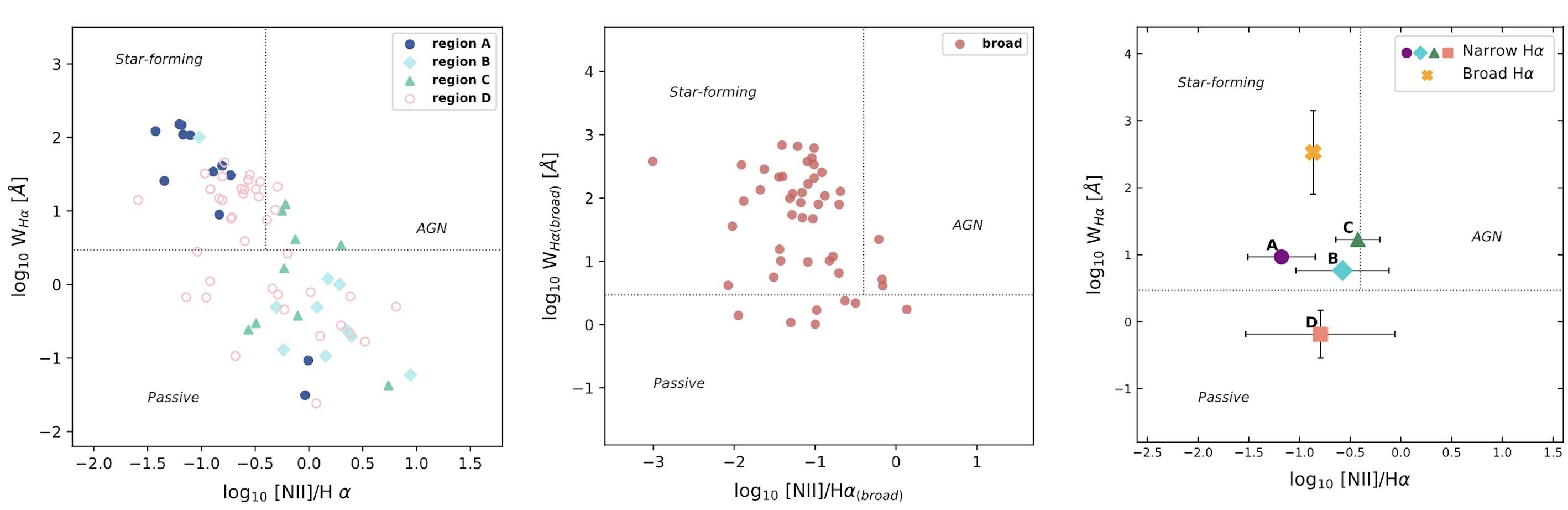}
\caption{WHAN analysis for ionized gas in 3C 297. Spaxel selection was done with SNR $\geq 2$ in [NII] line flux. \emph{(Left)} Region-wise mapping of narrow-line emission signal in a $3'' \times 3''$ around the target. \emph{(Middle)} Broad-line emission diagnostic from in a $2'' \times 2''$ region of spaxels in the outflow wing in Fig. \ref{fig:second}. \emph{(Right)} WHAN plot of emission signal derived from an integrated line profile over all spaxels in each region. Star-forming nebulae appear to be the dominant ionization source in the Northern arc as expected from its UV-bright HST detection, along with some contribution from the more evolved pAGB star population. But the WHAN analysis also favours stellar ionization power for the core $-$ in case of both the slower-moving gas in nuclear regions and the high-velocity gas flow $-$ over photoionization from the AGN.
\label{fig:whan}}
\end{figure*}

\subsection{What ionizes the emission-line gas in 3C 297?} \label{sec:ionize}

As the H$\beta$ line emission is yet to be measured for this galaxy, standard Baldwin-Phillips-Terevich (BPT) diagnostics (\citealt{1981PASP...93....5B, 1987ApJS...63..295V}) could not be used to determine the ionization source of the emission-line gasses in this case. We use the WHAN analysis developed by \cite{2011MNRAS.413.1687C} which utilizes the equivalent width of H$\alpha$ line (W(H$\alpha$)) versus the [N II]/H$\alpha$ ratio to identify the source of ionizing photons.  
\newline

The WHAN diagnostics for 3C 297 line emission in 3C 297 are shown in Figure \ref{fig:whan}. The vertical line at log[N II]/H$\alpha = -$0.40 corresponds to the SF/AGN division, designed to differentiate regions where star formation provides all ionizing photons from those where the quasar's harder ionizing spectrum is at work. The division at logW(H$\alpha$) = 0.5 \AA $ $ represents the distinction between emission-line galaxies and galaxies without emission lines in their spectra. The WHAN diagnostic deems galaxies to be lineless or ``passive'' if the equivalent widths of both H$\alpha$ and [N II] fall below 3 \AA. Thus, in spaxels with W(H$\alpha$) $<$3 \AA$ $ the emission will neither be AGN-related nor due to starburst activity. These regions most likely have an evolved stellar population as the dominant ionizing source, mainly post-asymptotic giant branch (pAGB) stars.

To best utilize the WHAN probe, we spatially map the [N II]/H$\alpha$ emission-line ratio with spaxel selection based on SNR $\geq$ 2 in the [N II] line. As shown in Figure \ref{fig:whan}, our analysis has three facets. The first one maps the narrow-line emitting gas region-wise to reveal the dominant ionizing mechanism in different regions of 3C 297. The region demarcation is as described in Sec \ref{broad}. In the second, we focus on the ionizing sources for the faster moving, broad line-emitting gas. In the third panel of Figure \ref{fig:whan}, we show the results obtained from co-adding emission signal in all spaxels of each region as well as from co-added spaxels containing broad-line emission.

\begin{figure}
\epsscale{0.9}
\plotone{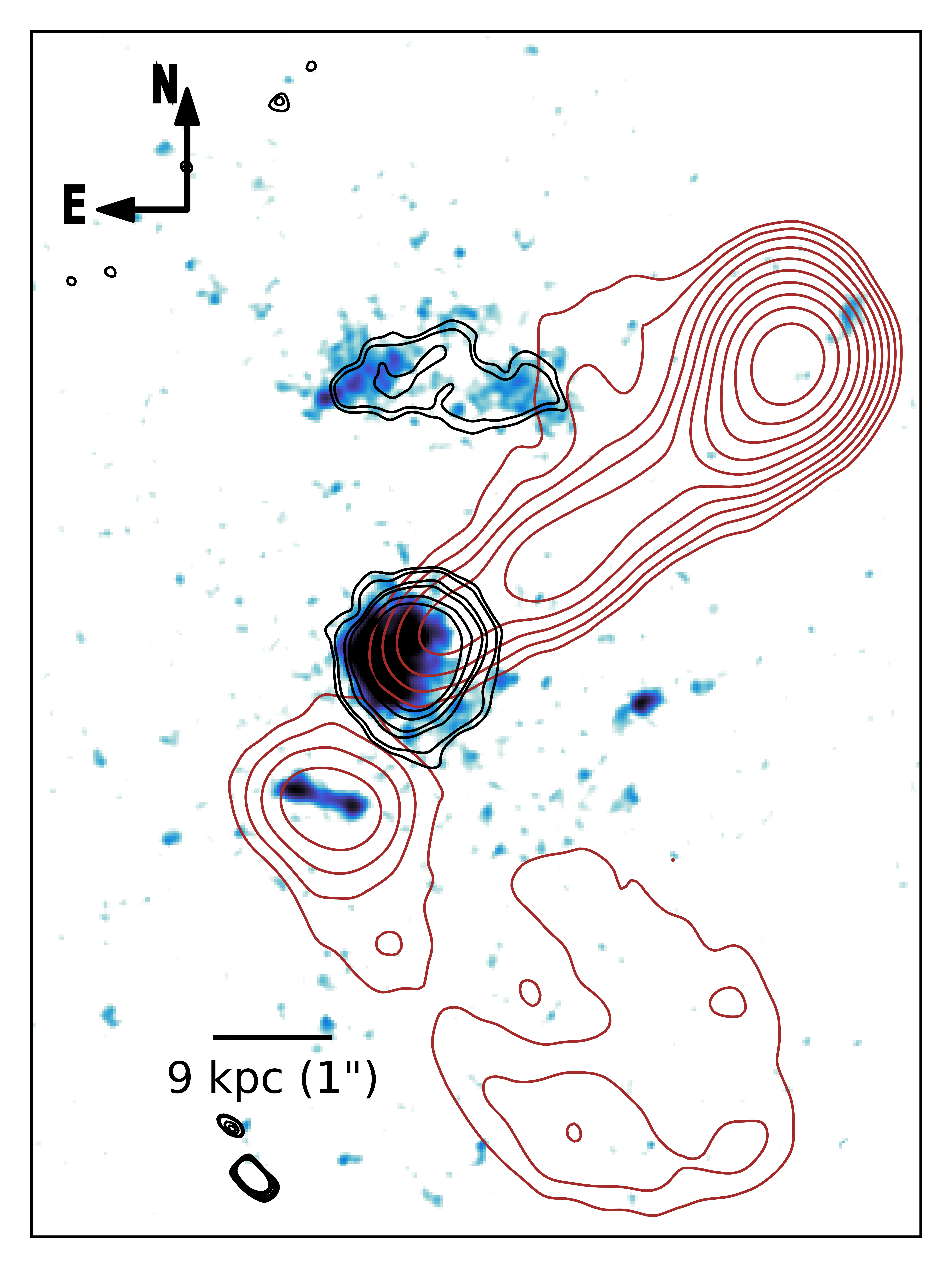}
\caption{HST/WFC3 F606W image of 3C 297 showing how the clumpy UV emission in the Northern arc and South-Western knots relates spatially with the jet lobe/hotspots. The black contours (levels $=[4, 5, 8, 10, 15, 20] \times 10^{-18}$ erg s$^{-1}$ cm$^{-2}$ spaxel$^{-1}$) trace the SINFONI narrow-H$\alpha$ emission, while maroon contours show the jet morphology from the VLA 8.4 GHz map. \label{fig:last}}
\end{figure}

Overall, the WHAN diagnostic shows moderate-to-low ionization over the $3''\times3''$ field of the 3C 297 system. The main source of ionizing photons is young, newly-formed stars combined with significant contribution from evolved pAGB star population, in both the core and the Northern arc. This is consistent with the bright, extended UV emission detected in the HST observation of 3C 297 (Figure \ref{fig:last}). The clumpy UV emission cospatial with narrow-H$\alpha$ emission in the arc (region D) and in circumnuclear region (region C) suggests recent or ongoing star formation. Photoionization due to the AGN may also be responsible for some of the ionized gas in the nuclear region, but most spaxels do not show sufficiently high [N II]/H$\alpha$ to indicate high-energy radiation from the AGN.  
\newline

One aspect that the WHAN diagnosis does not take into account is the shock-ionization from the action of the AGN jet. It is strongly expected, through findings from both simulations and observations of ISM-jet interaction, that shocks will have a significant contribution in the ionizing emission-line gas in jetted AGN, particularly when the radio source is still actively expanding through the ISM (e.g., \citealt{2022MNRAS.511.1622M, 2022MNRAS.516..766M, 2007ApJS..173...37S, 2002A&A...383...46M, 2000MNRAS.311...23B, 1996PASP..108..426M}). In fact, shocks associated with the radio source have been shown to dominate the ionization in some regions of the source such as those coincident with radio jet hotspots (e.g., Coma A; \citealt{2022A&A...657A.114C}). Collisional excitation via jet-driven shocks can easily produce the ionizing spectrum of hot young stars; indeed shocks of sufficient kinetic energy could mimic a higher, LINER-like ionization as well (e.g., \citealt{2018ApJ...864...90M, 2021MNRAS.504.5087M}). This is because at high shock velocities, emission from a photoionizing precursor can start to dominate the optical emission of the shock, resulting in a mixture of high- and low-ionization emission lines \citep{2008ApJS..178...20A}. 
  
The radio source in 3C 297 certainly appears to be actively pushing through the ISM and thus shocks are highly likely to be ionizing the gas in the vicinity of the jet. In the regions where narrow-line emitting gas is clearly disturbed by the jet (e.g., region B and D), shocks along with star-formation could be responsible for the ionizing radiation, while in the broad-dispersion outflow region (region C), shock-excitation may be the dominant mechanism given the spatial association of the outflowing gas with the jet lobes and hotspots. 
\newline

As a shock model diagnostic, the \textsc{mappings iii} (\citealt{2008ApJS..178...20A, 2013ascl.soft06008S}) line-ratio library has been used extensively in the literature and proven efficient in delineating the shock-ionization contribution in AGN hosts. We attempted to employ the shock, precursor, and shock+precursor models with solar abundance and a range of gas electron densities, against [N II]/H$\alpha$ and [S II]/H$\alpha$ ratios, to estimate possible shock velocities in 3C 297. However, the limited availability of diagnostic lines proved insufficient to draw conclusions from these models.

\subsection{What powers the outflow?} \label{sec:power}

The next step is to understand the physical mechanisms that can expel the ionised gas out to a distance of a few kpc from the center of the host galaxy. In 3C 297, we see a clear spatial correlation between the Southern jet and the outflowing gas (Figure \ref{fig:radio-overlays}, third panel). A comparison of the outflow and the radio jet energetics can inform us whether the outflow is being accelerated by the jet. But it is interesting to explore whether radiation pressure from the quasar  may be powerful enough to drive the outflow in this case. 
\newline

Using 1.4 GHz luminosity of 1.619 Jy for the 3C 297 radio source \citep{2015MNRAS.446.2985V}, we infer the kinetic luminosity $P_{jet}=1.76\times10^{46}$ erg sec$^{-1}$, employing the radio luminosity to jet power conversion relation from \cite{2010ApJ...720.1066C}. The bolometric power of the AGN radiative emission scales with the nuclear X-ray luminosity \citep{2020A&A...636A..73D}. For 3C 297, the X-ray luminosity is measured to be $1.15\times10^{44}$ erg sec$^{-1}$ \citep{2023ApJS..264....6M}, which translates to $L_{AGN}=1.35\times10^{45}$ erg sec$^{-1}$. 
\newline

For outflow energetics, we estimate the gas velocity using the following generally adopted relation (e.g., \citealt{2023A&A...678A.127V}), combining the centroid velocity $v$ and the FWHM width of the broad (outflow) component from the multi-Gaussian emission-line fitting:

\begin{equation}
v_{out} = v + \textrm{FWHM/2 = 1050 km s}^{-1}
\end{equation}

Taking the radius for the outflow from spatial projection as R = 8.6 kpc, the dynamical time is given by:

\begin{equation}
\textrm{t}\ = \textrm{R}/\textrm{v}_{out} = \textrm{8.13 Myr}
\end{equation} 

which is consistent with the expected typical AGN lifetime (e.g., \citealt{2001ApJ...547...12M}). 

The mass of ionized gas contained in the outflow is estimated by converting the extinction-corrected flux of the outflow component of H$\alpha$ (\citealt{2023A&A...678A.127V, 2017A&A...601A.143F, 2015A&A...580A.102C}) assuming electron temperature T$_e$ = $10^4$ K, as:

\begin{equation}
\textrm{M}_{out}/\textrm{M}_{\odot} = 0.6 \times 10^9\ \left(\frac{\textrm{L}_{H\alpha}}{10^{44}\ \textrm{erg s}^{-1}}\right) \left(\frac{500\ \textrm{cm}^{-3}}{n_e}\right)
\label{mass}
\end{equation} 

where the electron density $n_e$ in the outflow can be deduced from the [SII]$\lambda\lambda6716,30$ doublet ratio \citep{2006agna.book.....O}. 

Now, to compute the kinetic power of the outflowing gas, the geometry of the outflow region must be taken into account. The outflow is clearly directional (Fig. \ref{fig:second}), so we derive the physical properties of the outflowing gas from the observed line emissions adopting a simple conical (or bi-conical) outflow distribution uniformly filled with outflowing clouds.  

Given this hypothesis, the total mass outflow rate of a cone-shaped outflow out to a radius R for the ionized wind is:

\begin{equation}
\dot \textrm{M}_{out}/\textrm{M}_{\odot} = \textrm{v}_{out} \frac{\textrm{M}_{out}}{\textrm{R}_{out}} 
\end{equation} 

The outflow rate is thus independent of both the opening angle of the outflow and of the filling factor of the emitting clouds (under the assumption of clouds with the same density). 

Kinetic power can then simply be calculated as:

\begin{equation}
\dot \textrm{E}_{kin,out} = \frac{1}{2} \dot \textrm{M}_{out}\  \textrm{v}_{out}^2
\end{equation}

Estimation of the electron density $n_e$ is an important factor in this approximation. We use the integrated emission in regions A and B from the SINFONI S[II]-doublet flux map (Fig. \ref{fig:maps}), selecting only the spaxels with SNR $\geq2$, to derive $n_e \sim$30 cm$^{-3}$ in (region A) and $\sim$100 cm$^{-3}$ (region B). 
This produces a mass outflow rate in the range $2.5\times10^{3}-8\times10^{3}$ M$_\odot$ yr$^{-1}$ over these two regions (spanning a $\sim13$ kpc width centered on the nucleus.
So the kinetic power of the ionized outflow is in the range $3.2\times10^{45}-9.7\times10^{44}$ erg sec$^{-1}$. This would require a transfer of $\sim$5$\%-18\%$ of the jet kinetic power to the ISM. Transfer efficiencies of the jet energy to the kinetic energy of the ISM of $\leq30\%$ are expected from simulations (e.g.\citealt{2011ApJ...728...29W, 2016MNRAS.461..967M}). Therefore, the jet by itself is energetically able to drive the observed outflow. 

For the radiated energy of the AGN on the other hand, 
kinetic coupling efficiency needed to power the outflow is $\dot{E}_{kin}/L_{AGN}\sim$ 0.718, on the lower end of the outflow kinetic power. This is inconsistently high compared to theoretical predictions. Coupling efficiency within $\sim0.005–0.05$ is expected for an energy-conserving outflow,  (e.g. \citealt{2010MNRAS.401....7H, 2012ApJ...745L..34Z, 2014MNRAS.444.2355C, 2015ARA&A..53..115K}), or in the range $\sim0.001–0.01$ in case of a radiation-pressure driven outflow (e.g. \citealt{2018MNRAS.476..512I, 2018MNRAS.473.4197C}). This indicates that we are most likely underestimating the gas density $n_e$, and thus significantly overestimating the energy of the outflowing gas. The fraction of AGN’s radiative energy imparted to the accelerated gas falls within the expected range if a high-density outflow, with $n_e$ of the order of $\sim10^3$ cm$^{-3}$, is assumed.  

A higher-density consideration will in fact be reasonable in this scenario$-$ electron density in the warm ionized outflow phase is known to be highly uncertain in observationally-derived estimates \citep{2018NatAs...2..198H}.  Density diagnostics based on the [SII] and [OII] emission doublet ratios are only sensitive up to $n_e \sim 10^3.5$ cm$^{-3}$, and thus often result in $n_e$ estimations to be too low (e.g. \citealt{2018A&A...618A...6K, 2022ApJ...930...14R, 2023MNRAS.524..886H}). It is also expected that the density within the outflowing medium is non-uniform \citep{2018A&A...618A...6K}, varying radially around the nucleus, most likely due to the presence of shocks and high turbulence. As the derived outflow kinetic power is inversely proportional to the electron density, a locally underestimated $n_e$ can cause the resulting kinetic powers and coupling factors to be overestimated by multiple orders of magnitude.

\subsection{Star formation in 3C 297}\label{subsec:sfr}

\cite{2015A&A...575A..80P, 2016MNRAS.462.4183P} have analyzed the UV-to-IR spectral energy distribution of 3C 297, separating emission from the AGN torus, young (350 Myr) stars and older (900 Myr) stellar population. The IR luminosity attributed to star formation resulted in a star formation rate (SFR) of 160$\pm$0.2 M$_\odot$ yr$^{-1}$. The same study also computed the total stellar mass in 3C 297 host galaxy as $3\times10^{11}$ M$_\odot$.

As we derived in the preceding section, the rate of flow of the ionized gas in 3C 297 core region is of the order of $\sim$10$^3$ M$_\odot$ yr$^{-1}$. If indeed the bulk of the blue-shifted gas is outflowing, the efficiency of the removal of gas from the galaxy relative to the formation of stars could be quantified with the mass loading factor ($\eta$). In this case, $\eta$ is in the range 0.2$-$2 depending on the gas density. In other words, the rate of flow of gas mass due to the outflow would be 0.2$-$2 times the observed star formation rate. A low $\eta$ is consistent with 3C 297's relatively massive host galaxy, since high-mass galaxies are likely to retain more gas to form stars.
\newline

The HST F606 image revealed an interesting snapshot of extended UV emission in 3C 297 which in some regions aligns with the radio lobe and/or hotspots. Our SINFONI observation supports the  HST/UV picture of young star population distribution (Figure \ref{fig:last}). Bright narrow-line H$\alpha$ emission is detected in the Northern arc suggesting abundant star formation in the region. SINFONI H$\alpha$ map also shows bright emission from the gas clouds in the core, including narrow-H$\alpha$ co-spatial with the UV clumps extending outwards from the South-West of the core. This is the location of the blue-shifted gas flow, which is bordered on the North and South by the two jet lobes (region C;  Figs. \ref{fig:second} and \ref{fig:regions}). We do not detect H$\alpha$ from the Southern radio-aligned UV knots (Fig. \ref{fig:radio-overlays}). The sharp deflection of the jet from South-East to South-West at this location is a clear indicator of strong jet-ISM interaction and suggests that the UV knots are likely sites of jet-induced starbursts. The absence of H$\alpha$ detection in this region could be due to the star-forming nebulae consisting of young stars emitting in UV band but not being massive enough to generate enough ionizing photons to produce bright H$\alpha$ emission. 
It is also possible that star formation heated the surrounding clouds to X-ray emitting temperatures and the gas has not yet sufficiently cooled down to emit recombination H$\alpha$. 

In the Northern arc, the ongoing merger interaction in 3C 297 is likely to have ignited the starburst activity, which may have then enhanced due to the impact of the head of the jet advancing in the direction of the extended arc. The bright, energetic radio lobe enveloping the Western edge of the UV-emitting arc suggests vigorous interaction at the site. 

\begin{figure}
\epsscale{1.2}
\plotone{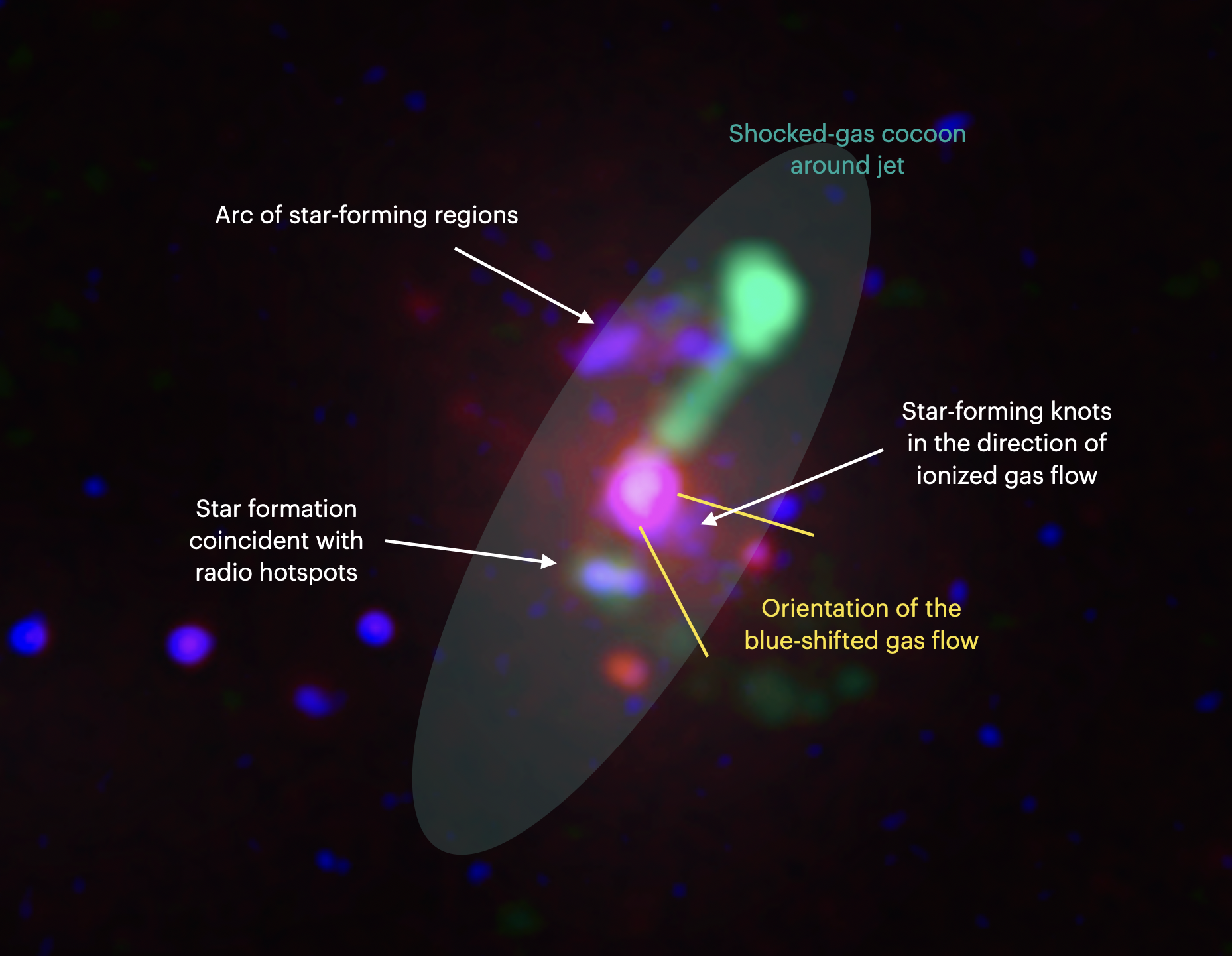}
\caption{Schematic view of the 3C 297 system drawn over a three-color composite image: HST F140W (red) for the rest-frame optical emission, HST F606W (blue) showing clumpy UV-emitting regions, and VLA/8GHz (green) radio image showing the jet morphology. The blue-shifted gaseous flow borders are marked with yellow lines and the locations of blue-excess from star-forming regions are highlighted with white arrows. North is up and East is left. \label{fig:model}}
\end{figure}

In the Southwest region, the narrow H$\alpha$ emission detected is blue-shifted, indicating bulk motion of ionized gas towards the observer. Despite this, the spatial association of UV-emitting clumps in this region with the outflow geometry is very interesting (see schematic model in Figure \ref{fig:model}). While star formation at the edges of an outflow cone has been observed (e.g., \citealt{2015ApJ...799...82C}), the outflow pressure is expected to drive dense ISM clouds directly in its path outwards, thereby suppressing star formation locally. In this case, a starburst may instead have been triggered due to outflow-induced compression. The resultant loss of momentum in the outflow would be consistent with the observed slowed down motion of the narrow-line gas in region C. However, since this region lies within the shocked ISM cocoon around the radio source (Figure \ref{fig:model}), it is thus more likely that interaction of the jet-driven shocks with a multi-phase ISM may have been responsible for both $-$ triggering star formation in the dense molecular clouds inside the cocoon as well as accelerating the ionized gas into a kpc-scale outflow. Shock-driven enhancement of star formation within the radio jet cocoon is expected (\citealt{1989MNRAS.239P...1R,1989ApJ...345L..21B, 2012MNRAS.425..438G, 2014ApJ...796..113D, 2018MNRAS.479.5544M}) and observed in radio galaxies (e.g., \citealt{2015A&A...574A..89S, 2022A&A...657A.114C, 2023MNRAS.519.3338T, 2024ApJ...965...17D}).

There is also a possibility that the blue-shifted line-emitting gas may be part of an inflow of material from surrounding medium into the merger, owing to its spatial offset from the jet axis and AGN ionization geometry.

\section{Concluding remarks} \label{sec:concl}

3C 297 is arguably a fascinating source and a potential probe of a variety of physical processes such as jet-ISM interaction, feedback from powerful radio sources and merger-induced fuelling and triggering of AGN, to name a few.
The integral-field data presented in this paper were limited to H$\alpha$, [N II] and [S II] lines, yet illustrate brilliantly the intricacy of ionized gas kinematics in this galaxy and the need for in-depth investigation by observing the emission-line regions with higher spatial resolution and at a broader range of optical/IR wavelengths.
\newline

The main results derived from our observations are summarized as follows:
\begin{itemize}
    \item The inner 15 kpc region exhibits fast-moving H$\alpha$ emitting gas. The broad-line map indicates rotating motions of ionized gas perpendicular to the North East-South West direction. 
    \item The Northern arc structure consists of mostly blue-shifted, slow-moving gas and is likely lined with ionized gas nebulae being powered by star formation resulting from merger-related interaction in the outer rim of the host galaxy, further enhanced by jet and/or bow shock impact. 
    \item The sharp deflection of the Southern jet at the radio-bright hotspots that coincide with UV knots suggests jet-induced star formation at the site of the hotspots. 
    \item A highly-directional, blue-shifted bulk motion of ionized gas is detected towards the South-West of the core. The spatial association of this $\sim10^3$ km s$^{-1}$ extended gas flow with the radio hotspots suggest a jet-driven outflow. But it is possible that the merger may be pulling in gas from the surrounding medium.
    \item Consistent with its Cosmic Noon environment, star-forming activity plays a major role in shaping the gas content in 3C 297. Most of the warm gas phase is being photoionized by hot, massive stars, with some contribution from collisional excitation by shocks driven by the expanding jet. 
    \item The powerful radio source appears to be the dominant driver of AGN feedback from the existing observational data for this system; likely causing the starburst-enhancing positive feedback along with the outflow of ionized gas out to $\sim$8 kpc in the galactic medium.
\end{itemize}

Dual-action feedback, i.e., radiation pressure from the accretion disk acting on the perturbed ISM together with the mechanical energy of the jet, may be influencing the gas kinematics in this system. Higher resolution [S II] map is needed for a more accurate estimation of the outflow electron density. Spatially-resolved observations of H$\beta$ and [O III] line-emission will aid in tracing the outflow geometry and improve ionization and kinematic analyses. These measurements are crucial for probing effects of the AGN accretion radiation. H$\beta$ observation for 3C 297 can also be used to construct an internal extinction map which will help study distribution of dust and improve UV light diagnostics.

Lastly, it will also be interesting to map the cold, molecular gas phase in 3C 297. Molecular gas measurements can provide insight into the ages and possible triggers of the starburst regions dispersed in the ISM of this merging system and shed light on the inward gas flows due to merger interaction that could trigger accretion onto the central black hole(s).


\begin{acknowledgments}

C.D. would like to extend thanks to Dr. Yjan Gordon for help with IFU data handling in the early stages of this project. C.D., C.O., and S.B. acknowledge support from the Natural Sciences and Engineering Research Council (NSERC) of Canada. 
Some of the data presented in this article were obtained from the Mikulski Archive for Space
Telescopes (MAST) at the Space Telescope Science Institute. These specific archival observations
utilized for anaylsis can be accessed via \dataset[doi: 10.17909/ea87-9v04]
{https://doi.org/10.17909/ea87-9v04}. This research has made use of NASA’s Astrophysics Data System Bibliographic Services, NASA/IPAC Extragalactic Database (NED), as well as the NASA/IPAC Infrared Science Archive (IRSA), which are funded by the National Aeronautics and Space Administration and operated by the California Institute of Technology. 
\end{acknowledgments}


\facilities{VLT, HST, VLA}


\software{Ned Wright's Cosmological Calculator \citep{2006PASP..118.1711W}, SAOImage ds9 \citep{2003ASPC..295..489J}, QFitsView \citep{2012ascl.soft10019O}, NumPy \citep{harris2020array}, Astropy (\citealt{2022ApJ...935..167A, 2018AJ....156..123A, 2013A&A...558A..33A}), Matplotlib \citep{Hunter:2007}, IPython \citep{2007CSE.....9c..21P}, \textsc{reproject} \citep{2020ascl.soft11023R}.}






\bibliography{ref_bib_list}{}
\bibliographystyle{aasjournal}



\end{document}